\documentclass[preprint,12pt]{elsarticle}

\usepackage{amssymb}
\usepackage{graphicx}
\usepackage{amsmath}

\begin{document}

\begin{frontmatter}

\title{Zigzag ordering, defects, and anomalous relaxation
	in antiferromagnetic Kuramoto lattices}

\author[1]{Priyanka D. Bhoyar} 
\affiliation[1]{organization={Department of Physics
	, Seth Kesarimal Porwal College
	}, addressline={Rashtrasant Tukadoji Maharaj Nagpur University}, 
       city={Kamptee, Nagpur},
       postcode={441001}, 
       state={Maharashta},
      country={India}}

\author[2]{Prashant M. Gade}
\affiliation[2]{organization={
	RJ College of Arts, Science and Commerce,},
          addressline={Ghatkopar West}, 
          city={Mumbai},
          postcode={400086}, 
          state={Maharashtra},
         country={India}}

\begin{abstract}
We investigate the nonequilibrium ordering dynamics of coupled Kuramoto oscillators with negative nearest-neighbor coupling, which induces a zigzag antiferromagnetic ordering. In one dimension, the defect density exhibits anomalously slow coarsening, decaying as $(D(t)\sim t^{-1/4})$ before saturating at a system-size-dependent time $(t_c(N)\sim N^z)$ with (z=2). The local persistence probability follows a stretched-exponential form, $(P(t)\sim \exp(-c t^\alpha))$, with $(\alpha=1/4)$. These exponents are observed  are independent of the magnitude of the coupling, which merely rescales the characteristic time scale. The equality $(\alpha=\delta=1/4)$ together with (z=2) is consistent with a distinct universality class.
These results demonstrate that deterministic nonlinear dynamics and geometric frustration alone are sufficient to generate slow relaxation and anomalous scaling, without quenched disorder or stochastic noise.

A continuum approximation and the corresponding coarse-grained partial differential equation provide a theoretical explanation for the observed anomalous exponents, while linear stability analysis accounts for the emergence of the zigzag ordered state. In two dimensions, geometric frustration inhibits complete ordering and gives rise to long-lived metastable domain-wall structures. An initial transient defect decay is observed before crossover and saturation. These results demonstrate how frustration and continuous phase variables can fundamentally modify coarsening dynamics and generate anomalously slow relaxation in deterministic many-body systems.
\end{abstract}

\begin{graphicalabstract}
\begin{figure}[h!]
    \centering
    \includegraphics[width=0.75\textwidth]{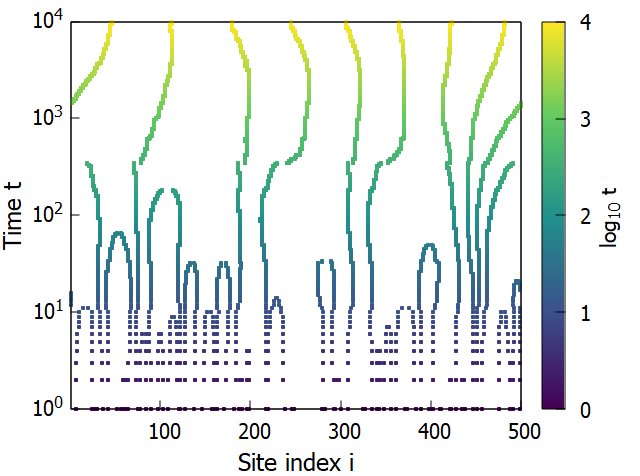}
    \caption*{The defect positions along a 1D Kuramoto chain are
 plotted as a function of time. Initially, the defects move and
 annihilate along constrained paths, but eventually settle into nearly equidistant configurations.}
\end{figure}
\end{graphicalabstract}

\begin{highlights}
 \item {We study antiferromagnetically coupled Kuramoto lattices that
 exhibit anomalous coarsening dynamics distinct from standard nonconserved phase ordering.}

 \item {In one dimension, the defect density decays as $D(t) \sim t^{-1/4}$ despite diffusion relaxation scaling with dynamical exponent $z=2$.}

 \item {Persistence follows a stretched-exponential form,
$P(t) \sim \exp(-c\, t^{1/4})$, with the stretching exponent equal to
 the defect decay exponent $(\alpha=\delta=1/4)$.}

 \item {Slow relaxation emerges purely 
from deterministic interactions,
without quenched disorder or stochastic noise.}

 \item {A continuum approximation and coarse-grained PDE explain the anomalous exponents, while linear stability analysis reveals the emergence of the zigzag ordered state.}

\end{highlights}


\end{frontmatter}



\section{Introduction}
The Kuramoto model~\cite{kuramoto1975international} is a popular and 
paradigm framework for studying synchronization
in the coupled oscillators. It shows how synchronization
can emerge in large populations of interacting oscillators
using a simple mathematical framework. It illustrates the 
transition from incoherent dynamics to synchronized behavior
and connects this transition to fundamental concepts in 
dynamical systems theory \cite{strogatz2000kuramoto}.
It is broadly applicable and easily
generalized, making it a valuable tool for studying
synchronization in physical, biological, and engineering 
systems \cite{acebron2005,Kuramoto1984,kozyreff2000global}.
Positive global coupling promotes collective synchronization
and can produce a synchronization transition even in populations
of nonidentical oscillators, a phenomenon that has been
extensively studied within the Kuramoto framework \cite{Kuramoto1975,acebron2005}.

Studies of negative (repulsive or antiferromagnetic) coupling in the Kuramoto model are comparatively less explored than the extensively studied attractive-coupling regime, despite their relevance to inhibitory neural interactions, Josephson-junction arrays, and active matter systems. In contrast to attractive coupling, which promotes phase synchronization, repulsive interactions favor phase differences of $\pi$, leading to antiphase ordering and the emergence of nontrivial spatial structures. Early work by Daido \cite{daido1992quasientrainment} demonstrated that frustrated oscillator interactions can give rise to slow relaxation and quasientrainment phenomena. Subsequently, Hong and Strogatz \cite{hong2011kuramoto,hong2012mean} investigated globally coupled oscillator populations with mixed attractive and repulsive interactions, revealing clustered and traveling-wave states. Antiphase ordering in one-dimensional repulsively coupled Kuramoto lattices was studied by Giver \textit{et al.} \cite{giver2011phase}, while coupled circle maps with negative coupling were shown to exhibit power-law persistence decay and traveling-wave structures \cite{gade2007power}. Repulsive coupling can also generate antiferromagnetic-like zigzag states in one dimension and checkerboard patterns in two-dimensional lattices \cite{giver2011phase,warambhe2023approach}. However, the nonequilibrium coarsening dynamics, defect kinetics, persistence, and scaling properties of Kuramoto lattices with purely antiferromagnetic interactions remain largely unexplored.



In this work, we investigate these transitions from a statistical physics perspective, focusing on defect dynamics and coarsening behavior. In oscillator systems, defects correspond to sites where the local phase deviates from the ideal ordering pattern, or equivalently, to domain walls separating ordered regions. Their dynamics provides valuable insight into the ordering process and the approach to the asymptotic state.

In ferromagnetic (FM) systems, interactions favor uniform alignment, and defects appear as domain walls or vortices that gradually annihilate as the system coarsens toward global order \cite{bray1994theory}. For systems in the Ising universality class, the defect density typically decays as a power law characterized by a dynamical exponent (z=2), while the persistence probability decays with exponent $\theta \approx 0.375$ \cite{bray1994theory,derrida1994non}. Similar coarsening behavior has also been reported in coupled map lattices exhibiting asymptotic antiferromagnetic order \cite{gade2013universal}.

However, not all ordering transitions belong to the Ising universality class. An analogue of the antiferromagnetic Kuramoto lattice is provided by coupled circle maps with negative coupling. These systems exhibit a power-law decay of persistence over a range of negative coupling strengths \cite{gade2007power}, despite the absence of complete asymptotic ordering. Thus, persistence scaling can arise even in dynamically active states. The present work demonstrates that a repulsively coupled Kuramoto chain exhibits slow defect-mediated relaxation toward antiferromagnetic order, with coarsening dynamics that differ from the standard Ising picture. We characterize the decay of defect density and persistence, derive a continuum description of the ordering process, and show that the resulting scaling behavior is governed by a distinct coarsening mechanism. In higher dimensions, geometric frustration can further influence the ordering dynamics and the nature of the asymptotic states.

In this work, we show that the 1D Kuramoto AFM system defies 
simple categorization, exhibiting unique scaling 
exponents that reveal a breakdown of standard dynamic 
scaling. We derive continuum differential equation for the same
and argue that 
the time decay exponent -$1/4$ follows from the biharmonic 
operator in the resultant equations. However, the dynamic 
exponent $z=2$, which makes the system anomalous. We also study
slope and field persistence in this system, which obeys stretched exponential decay, and we give an
argument for the same. We note that all these exponents do not 
depend on the values of the parameters as long as there is AFM 
interaction. 

\section{Model}
\subsection{Repulsively coupled Kuramoto lattice}
We consider a 1D chain of $N$ oscillators with periodic
boundary condition ($\phi_{N+1}(t)=\phi_1(t)$ and $\phi_0(t)=\phi_N(t)$) and
\emph{AFM coupling} $K<0$. Each oscillator 
$i$ is characterized by a phase $\phi_i(t)$.
The dynamics of the system is governed by the following
equation.
\begin{equation}
\dot{\phi}_i = \omega + K \big[\sin(\phi_{i+1}-\phi_i) + \sin(\phi_{i-1}-\phi_i)\big],
 \quad i=1,2,3,\dots,N,
\end{equation}
All oscillators share the same natural frequency $\omega$ and
interactions are restricted to nearest-neighbors along the
chain. For identical oscillators, the natural frequency $\omega$ can be eliminated by transforming to a rotating reference frame, $\theta_i=\phi_i-\omega t$. Furthermore, the magnitude of the coupling can be absorbed into a rescaling of time, $\tau=|K|t$, so that only the sign of $K$ affects the dynamics. In the present work, we therefore focus on the antiferromagnetic case $K<0$. 
We use $K$=-1.4 and $\omega$=0.1 for simulation. For $K<0$ 
the results remain qualitatively unchanged.
For numerical convenience, the phase variables are represented modulo $2\pi$ and are restricted to the interval $[-\pi,\pi)$. Since the coupling depends only on $\sin(\phi_j-\phi_i)$, so this representation does not alter the numerical integration of the oscillator dynamics.

In 2D, we consider an $N \times N$ lattice with nearest-neighbor couplings
. The dynamics of the system is governed by following equation.
\begin{equation}
\begin{split}
\dot{\phi}_{i,j}
=
\omega
+
K\left[
\sin\left(\phi_{i+1,j}-\phi_{i,j}\right)
+
\sin\left(\phi_{i-1,j}-\phi_{i,j}\right)
+\right.\\
\left.
\sin\left(\phi_{i,j+1}-\phi_{i,j}\right)
+
\sin\left(\phi_{i,j-1}-\phi_{i,j}\right)
\right]
\end{split}
\label{eq:2d_nn}
\end{equation}
where, $i=1,2,3,\dots,N$ and $ j=1,2,3,\dots,N.$

\subsection{Defect and order parameter definitions}
Defect Density $D(t)$:
Analogous to \cite{warambhe2023approach},
we define the local slope as
\begin{equation}
s_i = \mathrm{sign}\!\left(\phi_i - \phi_{i-1}\right),
\end{equation}
such that $s_i = 1$, if $\phi_i > \phi_{i-1}$, and $s_i = -1$, 
otherwise. A defect occurs at the site $i$ whenever the 
two consecutive local slopes have the same sign,
\begin{equation}
\vert s_i +s_{i+1} \vert= 2,
\end{equation}

This corresponds to a breakdown of the perfect antiferromagnetic
(zigzag) ordering \(\ldots,+,-,+,-,\ldots\).

Our order parameter is the normalized defect density:
\begin{equation}
    D(t) = \frac{1}{2N} \sum_{i=1}^{N} \vert s_i + s_{i+1} \vert.
\end{equation}

The equations of motion are integrated using a fourth-order
Runge--Kutta scheme with time-step $\Delta t = 0.01$, for
various system sizes $N$ and integration times $T$, with
averages taken over at least $N_c=400$ independent initial conditions.

We observe that the defect density decay as a power-law
with exponent -$1/4$ at early times. 
This decay, however, does not continue indefinitely and 
eventually saturates at a characteristic time $t_c$.
Importantly, the saturation value decreases
with increasing system size $N$, indicating that long-range 
order is recovered in the thermodynamic limit. A noteworthy aspect of the present results is that the system is deterministic, translationally invariant, and free of quenched disorder. Nevertheless, the defect density exhibits an unusually slow power-law decay, $D(t)\sim t^{-1/4}$, in contrast to the $t^{-1/2}$ behavior typically associated with curvature-driven coarsening in systems with dynamical exponent $z=2$. Thus, the anomalous coarsening observed here arises from the intrinsic dynamics of the antiferromagnetically coupled Kuramoto lattice rather than from randomness, noise, or static disorder.

\section{Results: 1D Coarsening}
\begin{figure}[h!]
    \centering
    \includegraphics[width=0.8\textwidth]{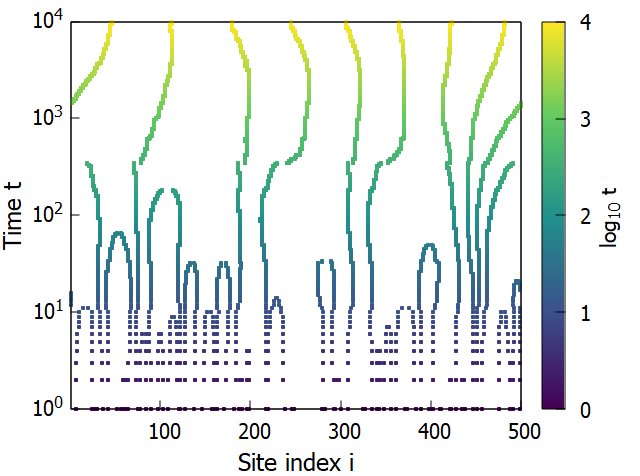}
    \caption{Scatter plot of defect positions along a 1D Kuramoto
 chain as a function of time. The y-axis is logarithmic to capture
 both early and late time dynamics. Defects initially move and
 annihilate along constrained paths, but eventually settle into
 nearly equidistant configurations.}
    \label{fig:defect_trajectories}
\end{figure}

Figure.\ref{fig:defect_trajectories} illustrates the
space-time trajectories of defects in the 1D Kuramoto
chain. At early times, defects move along constrained,
quasi-closed paths imposed by the underlying zigzag
ordering. Annihilation occurs primarily when such paths
intersect, leading to a slow decay of defect density, 
consistent with the observed $D(t) \sim t^{-1/4}$ behavior. 
Over time, the typical separation between defects increases,
which suppresses further mobility. Consequently, the system 
reaches a long-time configuration where defects become nearly 
equidistant and effectively frozen, resulting in a nonzero
residual defect density.

In the presence of a finite intrinsic frequency $\omega$,
the spatially averaged phase evolves as
$\frac{d}{dt} \langle \phi \rangle = \omega$,
leading to a uniform linear drift of the entire phase
field. Since the equations of motion depend only on phase
differences, this global rotation has no influence on the
internal dynamics, defect structure, or persistence 
properties of the system. To eliminate this trivial zero
mode and prevent unbounded growth of the phase variables, we
subtract the instantaneous spatial mean of the phase at each
time step for Figure.\ref{fig:defect_trajectories}. This
procedure corresponds to working in a rotating reference
frame and does not alter any physically relevant observables. 
The defects do not behave like annihilating random walkers.
Instead, they move in circular paths and annihilate each other
until the system reaches saturation.

\begin{figure}[htbp]
    \centering
    \includegraphics[width=0.65\linewidth]{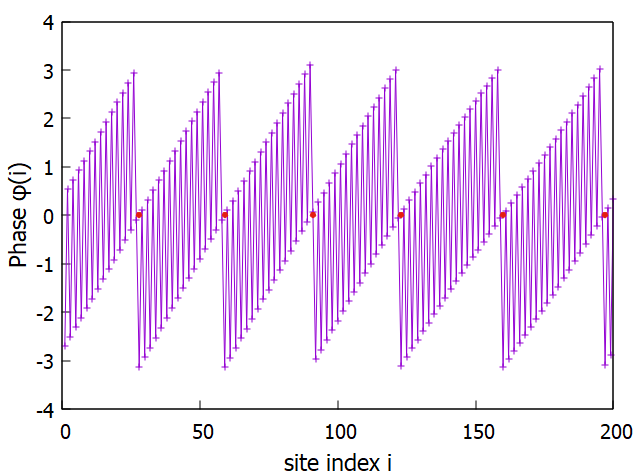}
    \caption{Asymptotic phase configuration for sites
    $i=1$--$200$
 in a 1D Kuramoto AFM chain. The purple points shows the phase
 profile $\phi(i)$, while red dots indicate the remaining defect sites.
 Most defects have annihilated, leaving a few residual defects that are roughly equidistant and pinned by the zigzag order.
 This illustrates the constrained mobility of defects at late times.}
    \label{fig:asymptotic_config}
\end{figure}

The asymptotic configuration of the 1D Kuramoto AFM chain is
shown in Figure.~\ref{fig:asymptotic_config}. After long time 
evolution, the oscillator phases arrange into an almost perfect
zigzag pattern. Most defects have annihilated, leaving a few
residual defects that are roughly equidistant and pinned by the
underlying order. The profile of the phase is plotted in
purple, while the residual defect sites are highlighted as red
dots. The spacing between asymptotic defects (if any) increases
as the size of the system increases.

At long times, the phase profile consists of extended zigzag 
domains separated by a small number of isolated defects 
(Figure.~\ref{fig:asymptotic_config}). Within each domain, the 
wrapped phase varies smoothly along the lattice, and the even
and odd sublattices form two slowly drifting branches of
opposite sign. The local zigzag structure is preserved
throughout these regions. Defects appear at isolated sites
where this alternating pattern breaks down. In the wrapped
representation, these sites correspond to locations where the
phase crosses zero ,while the neighboring sites lie close to 
$\pm $ $\pi$. Such configurations cannot be removed by a smooth
adjustment of the phase field and therefore represent genuine
phase-slip events separating zigzag domains of opposite parity.
As time progresses, the distance between defects increases, and
their mobility becomes strongly suppressed. The system then
approaches an asymptotically frozen state with a small but
finite residual defect density. This constrained dynamics is
responsible for the unusually slow decay of defects.

\subsection{Defect density and dynamical scaling}
\begin{figure}[htbp]
    \centering
    \includegraphics[width=0.65\linewidth]{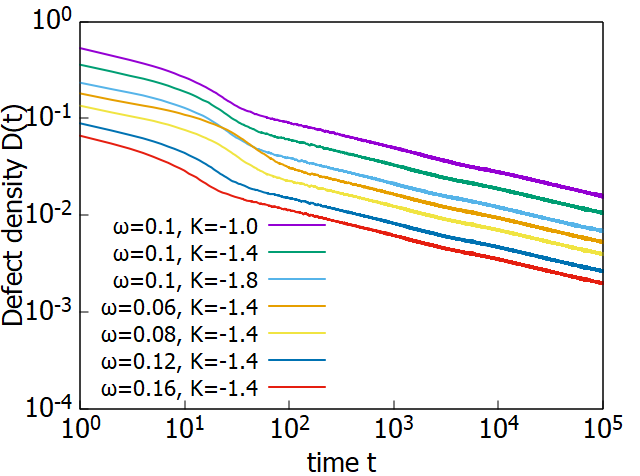}
    \caption{
Log-log plot of time evolution of the defect density $D(t)$ for different values of 
$\omega$ and $K$ in a 1D Kuramoto chain of size $N=50000$. 
After short transients, all curves collapse onto a universal 
power-law decay $D(t) \sim t^{-1/4}$, demonstrating parameter 
independent anomalous coarsening in the AFM 
Kuramoto chain. The y-axis is multiplied by arbitrary constants for clarity.
}
    \label{fig:2}
\end{figure}

Figure \ref{fig:2} shows the time evolution of 
the defect density $D(t)$ for 
different values of the intrinsic frequency
$\omega$ and coupling strength $K$ in 1D Kuramoto chain of size $N=50000$. 
Despite variations in $\omega$ and $K$, all curves 
collapse onto a 
common algebraic decay at late times, 
following $D(t) \sim t^{-1/4}$.
The observed 
parameter-independent scaling confirms that the slow relaxation 
is controlled by the underlying coarse-grained dynamics rather 
than by specific choices of $\omega$ or $K$ (as long as $K<0)$.

\begin{figure}[htbp]
    \centering
        \includegraphics[width=0.8\textwidth]{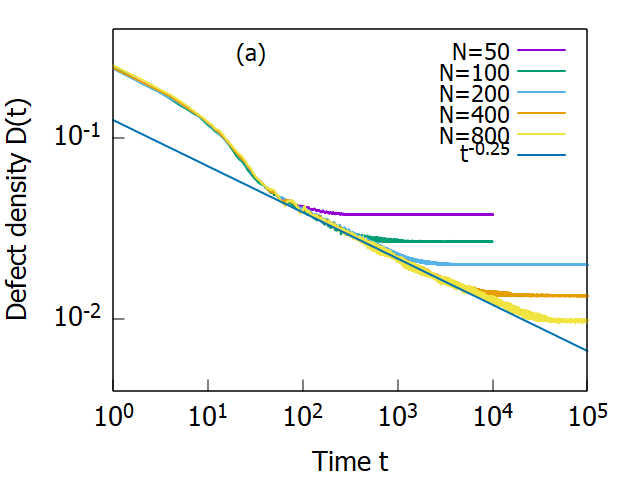}
         \includegraphics[width=0.8\textwidth]{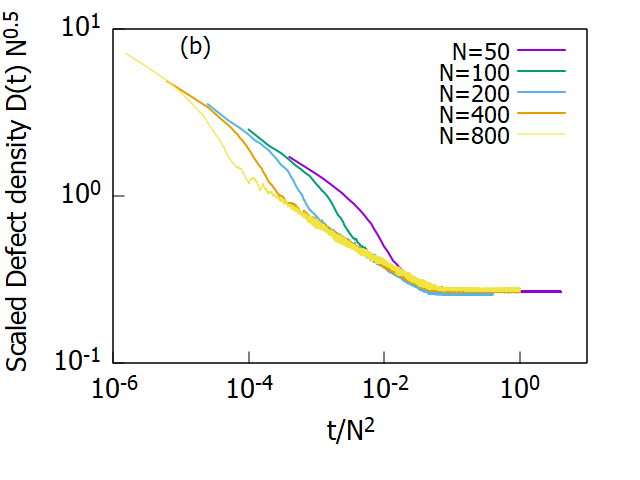}
    \caption{(a) Time evolution of defect density $D(t)$ for
 various system sizes $N=50,100,200,400,800$. The reference
 line shows $D(t) \sim \;t^{-1/4}$. (b) shows the finite-size
 scaling collapse of defect density: scaled time $t/N^2$ vs $D(t)N^{0.5}$ for 1D Kuramoto AFM chains. Each color corresponds to a different system size $N$.}
    \label{fig:defects_combined}
\end{figure}

Figure \ref{fig:defects_combined}(a) shows the time evolution
of the defect density $D(t)$ for different system sizes $N$=50,100,200,400,800. 
After an initial transient, the curves exhibit a slow algebraic
decay that is well described by $D(t) \sim  t^{-1/4}$,
as indicated by the reference line. The decay exponent appears 
largely independent of system size in the intermediate time 
regime, suggesting scale invariant coarsening dynamics before
finite-size saturation.

Dynamical exponent $z$:
The dynamical exponent $z$ is extracted using finite-size
scaling of the defect density. 
Assuming the scaling form
\begin{equation}
 \label{eq:fs}
D(t,N) \sim N^{-\delta z} f\!\left( \frac{t}{N^{z}} \right),
\end{equation}
Data for different system sizes \(N=50,100,200,400,800\) are 
collapsed onto a single universal curve by appropriately 
rescaling time as \(t/N^{z}\).

Figure~\ref{fig:defects_combined}(b) demonstrates finite-size
scaling of the defect density. When time is rescaled as
\(t/N^{2}\) and the defect density as \(D(t)N^{0.5}\), 
data for different system sizes collapse onto a single master
curve. This scaling collapse supports the dynamic scaling form
(Eqn.\eqref{eq:fs})
with dynamical 
exponent \(z \approx 2\) and decay exponent
\(\delta \approx 1/4\). The collapse confirms that 
the late time dynamics are governed by diffusive scaling
with a characteristic time scale proportional to \(N^{2}\).

Finite-size scaling confirms $z=2$, implying diffusive
domain growth $L(t) \sim t^{1/2}$.
This combination $(\delta=1/4, z=2)$ violates standard 
non-conserved coarsening scaling $(\delta=1/2, z=2)$, 
indicating \emph{anomalously slow defect annihilation}
despite diffusive domain growth. There is no crossover to higher
exponent. However, at $t\sim N^2$, the defect density 
is of the order $\sqrt{N}$ and defect density does not decrease 
any further. 

\subsection{Persistence}
Two types of persistence are defined.
(a) The slope persistence: $P_s(t)$ is the probability that
the local gradient variable
$s_i(t)=\mathrm{sign}(\phi_i-\phi_{i-1})$ has retained its
initial sign up to time $t$. (b) The field persistence: $P_\phi(t)$
is the probability that the phase variable $\phi_i(t)$ has not
crossed reference level $\phi=0$ up to time t.
Both $P_s(t)$ and $P_\phi(t)$ show stretched exponential decay:
$P(t) \sim \exp(-c \, t^\alpha), \quad \alpha = 1/4$,
saturating at finite $P_\text{sat} > 0$.  

\begin{figure}[htbp]
    \centering
    \includegraphics[width=0.65\linewidth]{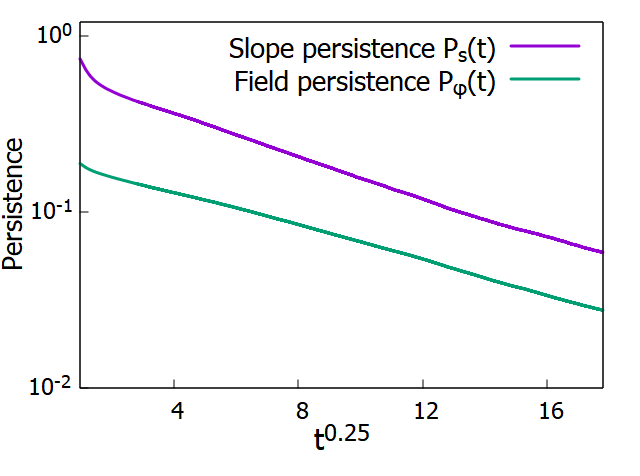}
    \caption{Persistence of the slope (purple) and field
 (green, scaled by 0.13 for clarity) in a 1D Kuramoto AFM chain of size $N=50000$. The horizontal axis is $t^{1/4}$, revealing a
 stretched exponential decay of persistence over time. Both field and slope persistence exhibit slow relaxation.}
    \label{fig:persistence}
\end{figure}

\begin{figure}[htbp]
    \centering
    \includegraphics[width=0.8\textwidth]{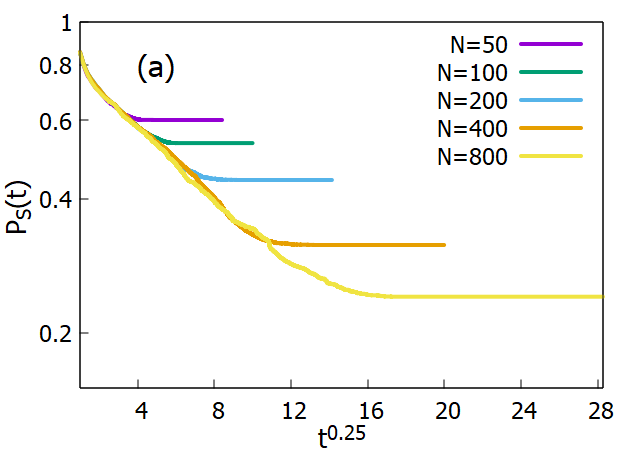}
    \includegraphics[width=0.8\textwidth]{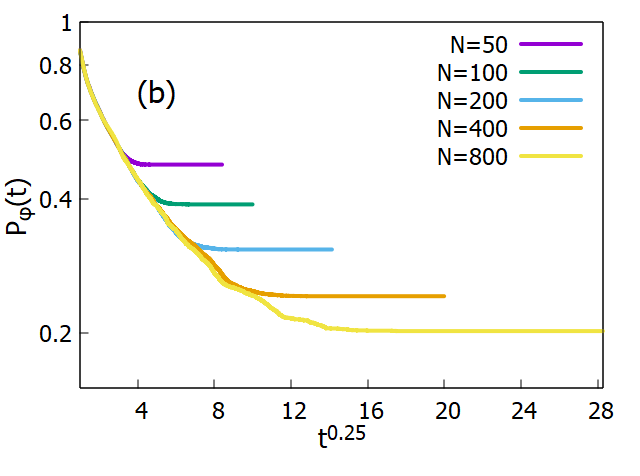}
    \caption{
    (a) shows the slope persistence probability $P_s(t)$ as a
    function of $t^{1/4}$. (b) shows the field persistence 
    probability $P_\phi(t)$ as a function of $t^{1/4}$ for
    system sizes $N=50,100,200,400,$ and $800$.
    At early times, persistence for all system sizes collapses 
    onto a straight line, demonstrating a stretched
    exponential decay $P(t)\sim \exp(-c t^{1/4})$ for both types of persistence. 
    Finite-size effects cut off the decay at a size-dependent 
    saturation time $t_{\mathrm{sat}}\sim N^2$, beyond which
    $P(t)$ approaches a nonzero
    saturation value.}
    \label{fig:persistence_t14}
\end{figure}

Figure~\ref{fig:persistence} shows the time evolution of 
persistence in a 1D Kuramoto chain of size 
$N=5\times 10^4$. Both  slope and field persistence 
are plotted as functions of $t^{0.25}$.
The data exhibits a stretched exponential decay, which is 
faster than a power-law decay but slower than a simple
exponential. In particular, the exponent  $\alpha$
of the stretched exponential coincides with the exponent 
$\delta$ observed in the decay of defect density 
$D(t)\sim 1/t^{\delta}$, providing a link between the two 
relaxation processes. This suggests that the slow dynamics of 
defects and the persistence of the phase field are controlled 
by the same underlying coarsening mechanism.

Figure \ref{fig:persistence_t14}(a) and (b) shows the persistence
 probabilities $P_s(t)$ and $P_\phi(t)$ 
 as a function of $t^{1/4}$.
For system sizes $N=50,100,200,400,$ and $800$, the data follow
a straight line at early times, confirming a stretched 
exponential decay $P(t)\sim \exp(-ct^{1/4})$.
The decay persists until a size-dependent cutoff time
$t_{\mathrm{sat}}\sim N^2$, beyond which $P_s(t)$ and $P_\phi(t)$ saturates
to a nonzero value. This saturation occurs at the same
time scale at which the defect density also becomes time 
independent, indicating that persistence is limited by
the finite-size relaxation of the underlying zigzag background.
Stretched exponential decay and finite saturation indicate 
kinetic trapping without quenched disorder, emerging from AFM 
frustration. Thus, while the early-time decay of
persistence is universal and size independent,
finite-size effects control the long-time behavior.

A clear qualitative distinction emerges between the
relaxational case $\omega=0$ and the driven case
$\omega \neq 0$. For $\omega=0$, both the field persistence
and the slope persistence approach finite asymptotic
values at long times for any finite lattice.
This saturation indicates that a nonzero fraction of
sites and bonds never change sign during the evolution.
Such behavior reflects that the defects annihilate
irreversibly, domain walls become pinned at large
separations, and the system approaches a metastable
frozen configuration that retains memory of the initial state.

In contrast, for finite intrinsic frequency ($\omega=0.1$),
both persistence decay rapidly to zero.
In this driven regime, local phase gradients continue
to fluctuate, and defects can be regenerated through
slow reversals of phase differences. As a result,
every site and bond eventually undergoes at least
one sign change, leading to complete loss of memory 
of the initial configuration.

Persistence has been widely employed as a probe of nonequilibrium relaxation in coupled dynamical systems and has proven particularly useful for identifying partially and completely arrested states. Depending on the nature of the dynamics, persistence may decay as a power law, characterized by a nontrivial persistence exponent \cite{mahajan2010transition}, or as a stretched exponential \cite{mahajan2018stretched}. The stretched-exponential decay observed in the present work, $P(t)\sim \exp(-c t^{1/4})$, is therefore consistent with the behavior of a broader class of coupled nonlinear systems that exhibit slow relaxation, constrained dynamics, and dynamical arrest.

\section{Continuum Limit and Band–Pass Instability}

We consider the nearest–neighbor Kuramoto dynamics
 with antiferromagnetic coupling $K<0$,
\begin{equation}
\dot{\phi}_i
=
K\left[
\sin(\phi_{i+1}-\phi_i)
+
\sin(\phi_{i-1}-\phi_i)
\right].
\end{equation}

Assuming a smoothly varying phase field $\phi_i \to \phi(x)$
 with lattice spacing set to unity, we perform a Taylor expansion about site $i$:
\begin{align}
\phi_{i\pm1}-\phi_i
&=
\pm \phi_x
+ \frac{1}{2}\phi_{xx}
\pm \frac{1}{6}\phi_{xxx}
+ \frac{1}{24}\phi_{xxxx}
+ \mathcal{O}(\partial_x^5).
\end{align}

Expanding the sine function and retaining terms 
consistent with the long–wavelength limit gives
\begin{align}
\sin(\phi_{i+1}-\phi_i)
+
\sin(\phi_{i-1}-\phi_i)
=
\phi_{xx}
+
\frac{1}{12}\phi_{xxxx}
+
\mathcal{O}(\partial_x^6),
\end{align}
where odd derivatives cancel by symmetry.
The continuum equation, therefore, becomes
\begin{equation}
\partial_t \phi
=
K\left(
\partial_x^2 \phi
+
\frac{1}{12}\partial_x^4 \phi
\right).
\end{equation}
The continuum expansion also generates nonlinear gradient terms, the leading one being
$-\frac{1}{6}\partial_x[(\partial_x\phi)^3]$.
For the zig-zag state, this term mainly renormalizes the coefficient of the $\phi_{xx}$ term after coarse-graining and does not alter the derivative structure of the continuum equation. We therefore expect it to modify only non-universal coefficients while leaving the asymptotic scaling unchanged. Thus the large-scale scaling behavior remains governed by the leading second, and fourth order derivative terms \cite{bray1994theory,cross1993pattern}.\\
Since $K<0$, we write
\begin{equation}
\partial_t \phi
=
-|K|\,\partial_x^2 \phi
-
\frac{|K|}{12}\,\partial_x^4 \phi .
\end{equation}

\subsection*{Linear stability}

For Fourier modes $\phi(x,t) \sim e^{ikx+\omega t}$, the dispersion relation is
$
\omega(k)
=
|K|k^2
-
\frac{|K|}{12}k^4 .
$
Long wavelengths ($k\to0$) are unstable,
$
\omega(k) \approx |K|k^2 > 0,
$
while short wavelengths are stabilized by the biharmonic term,
$
\omega(k) \sim -\frac{|K|}{12}k^4 < 0.$ 
The competition between the two terms produces a band-limited instability, with the fastest-growing mode obtained from $\frac{d\omega}{dk}=0$:
\begin{equation}
k^*=\sqrt{6},
\end{equation}
corresponding to a dynamically selected zigzag wavelength.
 Essentially, the corresponding wavelength $\lambda = 2$ is
 reasonably close to the continuum prediction $\lambda^* = 2\pi/k^* \approx 2.56$;
 the discrepancy is an artifact of the long-wavelength continuum approximation.

The early-time dynamics are governed by the biharmonic term, which produces 
surface-diffusion–type smoothing with amplitude decay $D(t)\sim t^{-1/4}$.
 In Fourier space, this corresponds to a dispersion relation $\omega(k)\sim -k^{4}$ 
dominating at intermediate wave numbers. At later times, the dynamics cross
 over to the long-wavelength sector $k\to 0$, where the quadratic contribution
 $\omega(k)\sim k^{2}$ becomes dominant. As a result, the characteristic
 relaxation time scales diffusively as $t_c\sim N^{2}$, implying a dynamical
 exponent $z=2$. Nonlinear mode coupling generates effective internal noise
 and produces finite roughness that saturates once the diffusive correlation
 length reaches the system size. The $k^{4}$ contribution primarily affects the relaxation of short, and intermediate-wavelength structures associated with local defects, while the $k^{2}$ term controls the longest-wavelength modes and hence the asymptotic finite-size relaxation time, $t_c \sim N^{2}$. The observed coexistence of $\delta = 1/4$ and $z = 2$ therefore suggests that defect decay and large-scale equilibration are governed by different parts of the spectrum.


\subsection*{Nonlinear saturation and coarsening}

Although the linearized equation contains a destabilizing negative 
diffusion term, the fundamental variable is an angle $\phi \in S^1$. 
The boundedness of the sine interaction prevents unbounded growth, 
and nonlinear effects saturate the instability.

The fourth–order term acts as a surface–diffusion regularizer. 
By analogy with Mullins-type surface diffusion dynamics, 
one expects a coarsening length scale $
L(t) \sim t^{1/4}$,
and a corresponding decay of gradients
$
D(t) \sim t^{-1/4}.$

However, instead of pure surface–diffusion scaling 
($z=4$), the system crosses over to effective diffusive coarsening,
$L(t) \sim t^{1/2}$,
while retaining an anomalous amplitude decay exponent
$\delta = \frac{1}{4}.$
We emphasize that the continuum description is based on a long-wavelength approximation and neglects higher-order nonlinear gradient terms. Although such terms are not expected to modify the observed scaling in the present one-dimensional system, they may influence the asymptotic behavior at very long times or in higher dimensions.

\subsection{Comparison with Other Models}
While $\nabla^4$ stabilization appears across pattern-forming 
PDEs 
(Kuramoto-Sivashinsky\cite{kuramoto1976persistent}, Swift-Hohenberg\cite{swift1977hydrodynamic}), clean $t^{-1/4}$ 
coarsening is the signature of Mullins-Herring surface diffusion. 
In these models of coarsening, the lateral length scale \(L(t)\) 
grows with time \(t\) according to a power-law: 
   $ L(t) \sim t^{1/4}$,

\subsubsection{Mullins--Herring Equation}

Mullins--Herring equation describes surface smoothing via 
adatom diffusion on solid surfaces.
\begin{equation}
    u_t = - u_{xxxx},
\end{equation}
linear theory predicts $ L(t) \sim t^{1/4},$ $\&\; z = 4.
$
\cite{mullins1957theory,mullins1959flattening,herring1950effect}

\subsection{Linear Relaxation of Corrugated Surfaces}

In the linear approximation, the relaxation of a corrugated 
surface back to equilibrium can be described by
\begin{equation}
\partial_t h = \nu \nabla^2 h - K \nabla^4 h,
\end{equation}
where the first term corresponds to evaporation 
(mass removal proportional to local curvature) and the second term 
corresponds to surface diffusion. 

Asymptotically, this model leads to a decay of surface roughness
 as $t^{-1/4}$, since evaporation removes mass uniformly while 
diffusion dominates the relaxation of long-wavelength 
fluctuations, effectively allowing the growth of large-scale domains.
This behavior has been discussed in detail by Villain \cite{villain1991continuum}.
   
 \subsubsection{Deterministic Coarsening of Mounds (\(\alpha\)-dependent slope steepening)}
The governing equation is given by 
\begin{equation}
\partial_t z(x,t)
=
- \partial_x^4 z
- \partial_x \left[
\frac{\partial_x z}{\left(1 + (\partial_x z)^2\right)^{\alpha}}
\right]
+ \eta(x,t).
\end{equation}

In the presence of noise, the typical mould size 
$N(t)\sim t^{1/4}$ always. For $\eta(x,t)=0$, $N(t)\sim t^{1/4}$
for $1\le \alpha \le 2$ \cite{torcini2002coarsening}.

We note that higher-dimensional models can also exhibit
 $t^{-1/4}$ decay\cite{bray1994theory}. 
However, for clarity and brevity, we restrict the present discussion to 
representative one-dimensional models that are most directly relevant to our system.

\section{Kinetic derivation of stretched exponential persistence}

The slope persistence probability $P(t)$ is defined as the 
probability that the slope variable
 $s_i(t) = \text{sign}(\phi_i(t) - \phi_{i-1}(t))$
 at site $i$ has never flipped from its initial value up to time $t$.
 Our numerical results indicate a stretched-exponential decay,
 $P_S(t) \sim \exp(-c t^{1/4})$, which we derive here by 
 linking the defect density $D(t)$ to the diffusive growth of
 the zigzag background.

\subsection{Defect Encounter Rate}
We assume that a flip in the local slope $s_i$ occurs only
when a defect passes through site $i$. In 1D Kuramoto 
AFM chain, defects are not independent random walkers; rather,
their motion is slaved to the equilibration of the underlying 
zigzag background. The characteristic length scale of these 
ordered domains grows diffusively as $\xi(t) \sim t^{1/2}$ 
(with dynamical exponent $z=2$).

The number of active defects within a single domain of size $\xi(t)$ is given by:
\begin{equation}
n_d(t) = D(t) \cdot \xi(t) \sim t^{-1/4} \cdot t^{1/2} = t^{1/4}.
\end{equation}

The effective encounter rate $\Gamma(t)$, or the flip rate per
 site due to defect passage, is the ratio of active defect count per unit time
\begin{equation}
\Gamma(t)  \sim \frac{t^{1/4}}{t} = t^{-3/4}.
\end{equation}

\subsection{Cumulative Hazard and Persistence}
Persistence $P(t)$ is the probability of survival under the 
cumulative hazard of encounters with defects. Integrating the rate
 $\Gamma(t)$ over time yields the total probability of a flip event:
\begin{equation}
\int_0^t \Gamma(t') \, dt' \sim \int_0^t (t')^{-3/4} \, dt' \propto t^{1/4}.
\end{equation}

Substituting this into the survival function\cite{klein2003survival}, we obtain:
\begin{equation}
P(t) = \exp\left( -\int_0^t \Gamma(t') \, dt' \right) \sim \exp(-c t^{1/4}).
\end{equation}
This derivation establishes that the persistence exponent $\alpha = 1/4$ 
is exactly equal to the defect decay exponent $\delta = 1/4$.

\subsection{Saturation and Finite-Size Effects}
This relaxation process continues until the correlation 
length $\xi(t)$ reaches the system size $N$ at the saturation time 
$t_c(N) \sim N^2$ At this point, the zigzag background is essentially 
frozen, the encounter rate $\Gamma(t) \to 0$, and the persistence 
reaches a nonzero plateau $P_{\text{sat}} > 0$. The equality
 $\alpha = \delta = 1/4$ characterizes a universality class 
defined by quartic relaxation occurring on a background that evolves via $z=2$ diffusion.

\section{Results: 2D Dynamics}
\subsection{Nearest-Neighbor Coupling ($K<0$)}
We study the deterministic dynamics of a two–dimensional 
lattice of phase oscillators with nearest–neighbor coupling. The evolution equation is
\begin{equation}
\dot{\phi}_{i,j}
=
\omega
+
K \sum_{\langle i',j' \rangle}
\sin\!\left(\phi_{i',j'} - \phi_{i,j}\right),
\end{equation}
where $(i,j)$ label lattice sites on a square lattice of size $N \times N$
 with periodic boundary conditions, $\omega$ is a uniform intrinsic frequency,
 and $K$ is the coupling strength. The summation runs over the four nearest neighbors.
 The equations of motion are integrated using a fourth–order Runge–Kutta (RK4) 
scheme with fixed time step $h=0.01$. 
Initial conditions are chosen independently at each lattice site from a 
uniform distribution in $(0,2\pi]$. The phase variables are represented modulo $2\pi$ and are restricted to the interval $[-\pi,\pi)$. All simulations are performed for a 
total integration time $T = N_T h$, where $N_T=10^4$.

\begin{figure}
    \centering
    \includegraphics[width=0.8\linewidth]{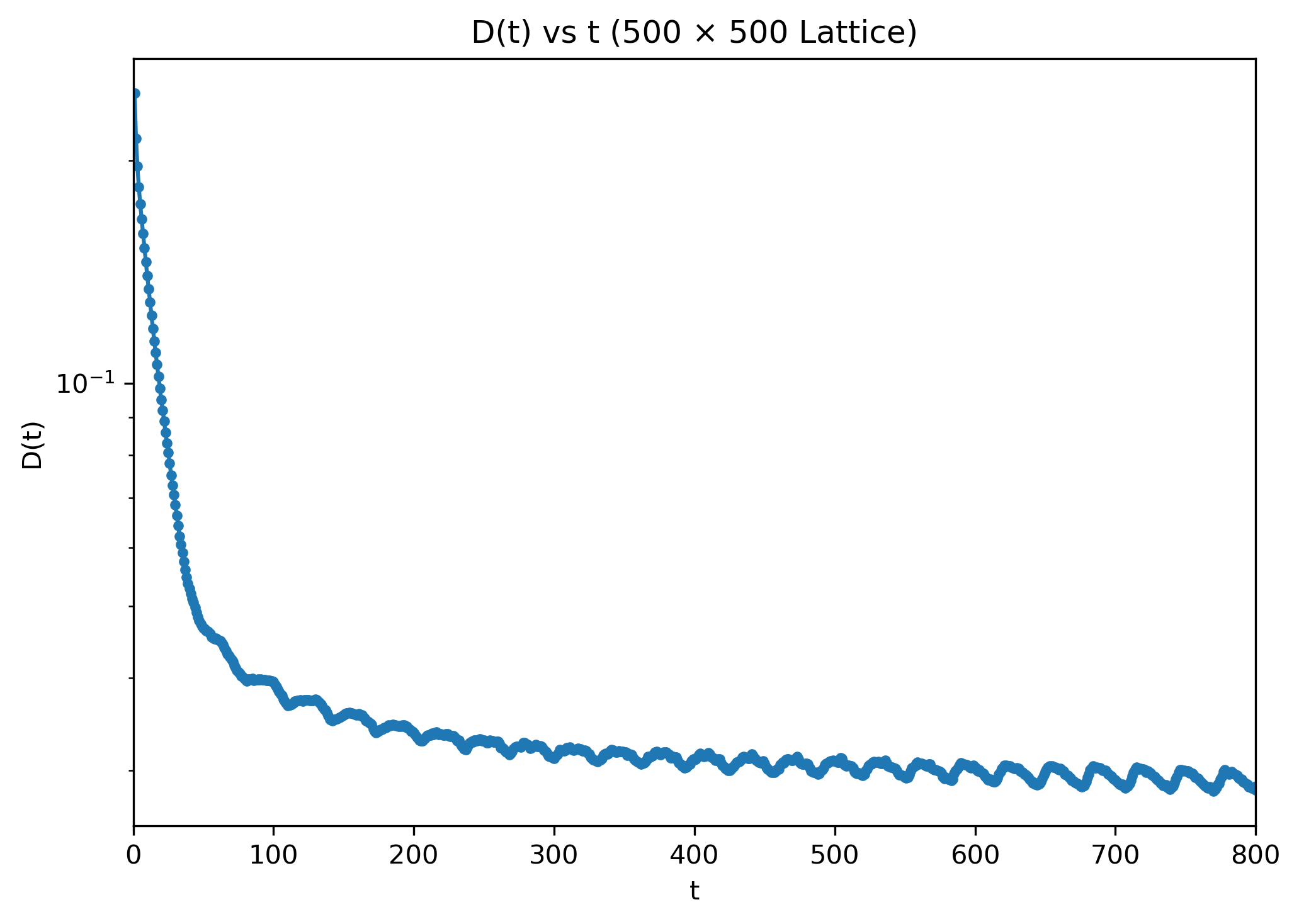}
    \caption{Defect density $D(t)$ plotted as a function of time $t$ for a $500 \times 500$ lattice. After an initial exponential decay, the defect density saturates at long times.}
    \label{fig:2d1}
\end{figure}

We compute a local slope-alignment order parameter based on the signs of discrete phase differences along both lattice directions. The measure quantifies the local alignment of phase gradients and provides an estimate of the density of domain-wall defects.
At early times, the defect density exhibits an approximate transient decay. However, this regime persists only over a limited temporal
window before crossing over to saturation.
At long times, geometrical frustration traps the system in 
metastable configurations, causing the defect density to 
saturate instead of decaying to zero(See Fig.\ref{fig:2d1})

\begin{figure}[htbp]


    \centering
    \includegraphics[width=\linewidth]{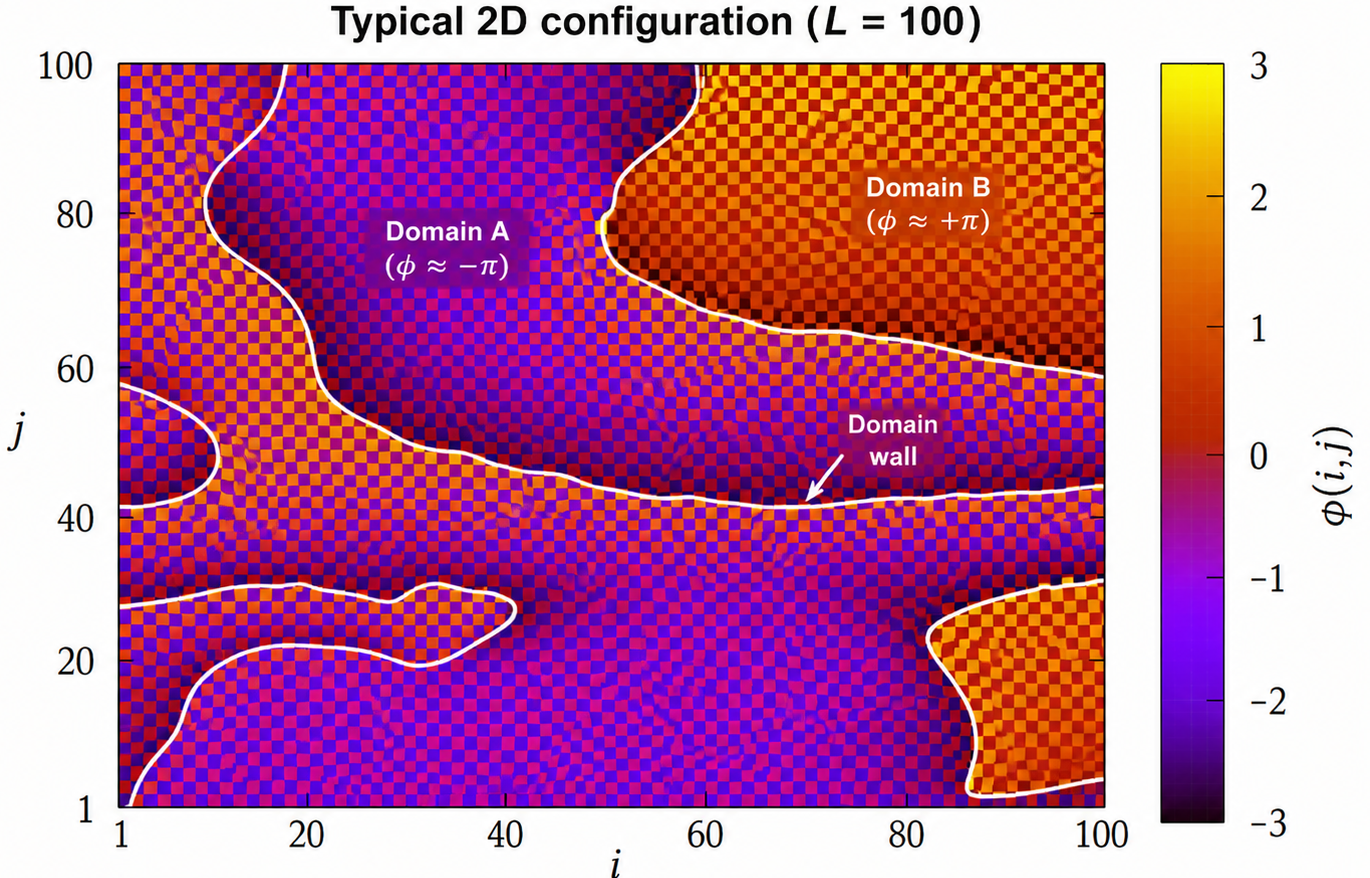}

\caption{Late-time phase configuration $\phi(i,j)$ for $K<0$. 
Late-time phase configuration exhibiting local checkerboard ordering and extended defect lines that prevent complete long-range order.}
\label{fig:spa2d_nm}
\end{figure}
Fig.\ref{fig:spa2d_nm} shows a representative phase
configuration for $K=-1.4$.  The fine checkerboard pattern
reflects strong local antiferromagnetic correlations, with
neighboring sites differing approximately by $\pi$. However,
the system does not develop a global staggered order. Instead, 
extended defect lines separate antiferromagnetic domains
with different staggered orientations. These domain walls
correspond to phase-slip 
regions where the local ordering pattern shifts, preventing the 
emergence of clear global staggered order
within the simulated time scales and system sizes.
The qualitative difference between one and two dimensions arises from the geometry of the defects. While one-dimensional defects are point-like and eventually annihilate upon encounter, two-dimensional defects organize into extended domain walls \cite{bray1994theory}. The elimination of such walls requires coordinated motion and shrinkage of entire line segments rather than local pairwise annihilation, substantially slowing the coarsening process and hindering the development of long-range order. This role of domain-wall excitations in frustrated XY-like systems has been widely discussed in the literature \cite{Denniston1997,Mouritsen1988}.
The resulting
state is therefore a frustrated antiferromagnetic
domain structure characterized by 
persistent interfaces rather than complete ordering.
The present 2D analysis is primarily qualitative.
A detailed theoretical understanding of the crossover from
the transient early-time relaxation to the saturated
frustrated state remains an open problem.


\section{Discussion and Conclusion}
We have shown that antiferromagnetically coupled Kuramoto 
lattices exhibit anomalous coarsening dynamics that differ 
fundamentally from standard nonconserved phase ordering. 
The observed slow relaxation and anomalous scaling emerge
entirely from deterministic nonlinear dynamics, while geometric
frustration further shapes the ordering process in higher dimensions,
without any need for quenched disorder or stochastic noise.
Furthermore, unlike spin systems where 
discrete variables are typically studied, here the dynamical 
degrees of freedom are continuous phase variables, and the 
defects are defined in terms of phase differences wrapped 
within $(-\pi,\pi]$, giving rise to nontrivial scaling behavior.

In one dimension, the system rapidly develops local zigzag 
order, but defect elimination proceeds unusually slow. The
defect density decays algebraically as $D(t) \sim t^{-1/4}$,
while the characteristic relaxation time scales diffusively as 
$t_c \sim N^2$, corresponding to a dynamical exponent $z=2$.
Such a value of $z$ is not unusual. Diffusive scaling with 
$z=2$ is a common feature of many relaxation systems, including 
the Edwards--Wilkinson equation\cite{barabasi1995fractal},
model A dynamics, and the long wavelength limit of the Cahn-
Hilliard equation\cite{bray1994theory}, where the Laplacian 
term $\nabla^2$ governs the asymptotic behavior. In the present 
system as well, the long wavelength $\nabla^2$ contribution 
dominates at late times, leading to diffusive scaling of the 
equilibration time. The novelty therefore does not lie in the value z=2 itself, but in its coexistence with the unusually slow defect-decay exponent $\delta=1/4$.
In standard nonconserved coarsening with $z=2$, one typically
expects $\delta=1/2$, corresponding to annihilating random 
walker behavior of defects. Here, the defect density decreases
much more slowly. The early time defect dynamics are therefore
fundamentally different from simple annihilating random walkers
and remain strongly constrained by the evolving zigzag 
background.

Persistence further highlights the unconventional nature of the
relaxation process. Both slope and field persistence follow a
stretched-exponential form, $P(t) \sim \exp(-c\, t^{1/4})$,
with a stretching exponent equal to the defect decay exponent.
 The equality $\alpha=\delta=1/4$ indicates that local memory 
loss and global defect annihilation are governed by the same 
underlying time-dependent mechanism. In the purely relaxation 
case ($\omega=0$), persistence saturates at long times for finite
 system sizes. This saturation coincides with the arrest of defect
 motion. Once the defect density reaches its finite-size plateau, the defects become effectively immobilized, and no further sign change
 occurs in the slope or field variables. As a result, persistence
 stops decaying because no additional defect passages take place.
 The observed plateau therefore, reflects finite-size dynamical freezing.
 In the thermodynamic limit, where the saturation time diverges
 as $t_c \sim N^2$, persistence would continue to decay over 
progressively longer time scales.

The continuum description clarifies the origin of these anomalous
 exponents. A long wavelength expansion produces a competition
 between an unstable second-order term and a stabilizing fourth-order
 term. While linear theory suggests surface-diffusion-type scaling, 
Nonlinear saturation and the compact $S^1$ topology of the phase variable
 renormalize the late time behavior. The zigzag background equilibrates
 diffusively ($z=2$), but defect motion remains slaved to a slower 
smoothing process. The resulting interplay yields diffusive length growth together with quartic amplitude decay, providing a theoretical explanation for the observed exponent $\delta$=1/4 within the continuum approximation.


In two dimensions, nearest-neighbor antiferromagnetic
coupling on 
the square lattice generates geometric frustration. Although strong 
local checkerboard correlations develop, staggered domains cannot align 
consistently across the entire system. Extended domain walls persist over the accessible simulation times and strongly inhibit the development of long-range order. At early times, the defect density exhibits a transient decay before crossing over to saturation. At long times, geometrical frustration traps the system in metastable configurations, preventing complete defect annihilation and leading to a nonzero saturated defect density.
Unlike the one-dimensional case, where defects are point-like and can eventually disappear through pairwise encounters, two-dimensional defects form extended interfaces whose removal requires collective rearrangements over large spatial scales. Consequently, geometrical frustration strongly suppresses the development of long-range order and traps the system in metastable antiferromagnetic domain configurations characterized by persistent domain walls.

\section*{AI-Assisted Writing Statement}
Generative AI tools were used only to improve the language and 
clarity of this manuscript. AI is used to write text in Fig. 9. The authors carefully reviewed and 
edited the output and take full responsibility for the final content.
\section*{Acknowledgment}
PMG thanks IMSc, Chennai, for hosting a visit and Prof. Sitabhra 
Sinha for discussions. PDB thanks Rashtrasant Tukadoji Maharaj 
Nagpur University for providing financial assistance (RTMNU/RDC/2024/242).


\end{document}